\documentclass[10pt,preprint] {aastex}

\usepackage{graphicx,epsfig,natbib,color}
\usepackage{lscape}
\usepackage{graphicx}
\usepackage{epsfig}
\usepackage{natbib}

\begin{document}
\title{Dynamic Evolution of an X-shaped Structure above a Trans-equatorial Quadrupole Solar Active Region Group}

\author{J. Q. Sun\altaffilmark{1,2}, X. Cheng\altaffilmark{1,2}, Y. Guo\altaffilmark{1,2}, M. D. Ding\altaffilmark{1,2}, and Y. Li\altaffilmark{1,2}}

\affil{$^1$ School of Astronomy and Space Science, Nanjing University, Nanjing 210093, China}\email{dmd@nju.edu.cn; xincheng@nju.edu.cn}
\affil{$^2$ Key Laboratory for Modern Astronomy and Astrophysics (Nanjing University), Ministry of Education, Nanjing 210093, China}

\begin{abstract}
In the solar corona, magnetic reconnection usually takes place at the singular configuration of magnetic field, in particular near a magnetic null owing to its high susceptibility to perturbations. In this Letter, we report a rare X-shaped structure, encompassing a magnetic null, above a trans-equatorial quadrupole active region group that is well observed by the Atmospheric Imaging Assembly (AIA). The observations show that this X-shaped structure is visible in all AIA EUV passbands and stably exists for days. However, possibly induced by flare activities at the northern part of the quadrupole active region group, the X-shaped structure starts to destabilize and meanwhile a jet erupted near its center at $\sim$ 15:05 UT on 2013 October 7. Through the non-linear force-free field modeling, we identify a magnetic null, which is above the quadrupole polarities and well corresponds to the X-shaped structure. After the jet eruption, the temperature and emission measure of the plasma near the X-shaped structure rise from $\sim$ 2.3 MK and $\sim$ $1.2\times10^{27}$ $\mathrm{cm}^{-5}$ at 15:01 UT to $\sim$ 5.4 MK and $\sim$ $3.7\times10^{27}$ $\mathrm{cm}^{-5}$ at 15:36 UT, respectively, revealed by the differential emission measure analysis, indicating that magnetic reconnection most likely takes place there to heat the plasma. Moreover, the height of the null has an increase of $\sim$ 10 Mm, which is most likely due to the partial opening of the field lines near the fan surface that makes the null underneath rise to seek for a new equilibrium.
 \end{abstract}

Online-only material: animations, color figures

\section{INTRODUCTION} 
Magnetic reconnection is generally believed to play an important role in converting magnetic energy to thermal, non-thermal, and kinematical energies to support various eruptive activities such as solar flares \citep[e.g.,][]{Luoni2007a, Masson2009a}, jets \citep[e.g.,][]{SunXuDong2012a, Filippov2013a}, and coronal mass ejections \citep[e.g.,][]{Ugarte-Urra2007a, Barnes2007a}. In the past, basic two-dimensional theoretical models of magnetic reconnection have been proposed, like the Sweet-Parker model \citep{Sweet1958a, Parker1957a} and the Petschek model \citep{Petschek1964a}. More complex three-dimensional situation has also been investigated both in theories \citep[e.g.,][]{Priest1996a} and simulations \citep[e.g.,][]{Galsgaard1997a, Galsgaard2005a}. However, the detailed physical processes involved in magnetic reconnection are still inconclusive and await more investigations, especially through high resolution observations.\par

One of the most possible sites for magnetic energy release is magnetic null, the magnetic field near which has a high susceptibility to perturbations \citep{Priest2009a}. It is found that magnetic nulls are widespread in the corona \citep{Regnier2008a, Longcope2009a} and may serve as a significant factor for coronal heating \citep{Priest2005a}. In 2D reconnection models \citep{Sweet1958a, Parker1957a, Petschek1964a}, magnetic null is always believed to be a site where anti-parallel magnetic fields dissipate. In 3D models \citep{Priest1996a, craig1996a,Priest2009a}, however, reconnection can also take place in spine and fan structures \citep{Lau1990a, Priest1996a}, which are ideal sites for current formation, besides magnetic nulls. \par

The direct way to seek the null is to perform an extrapolation of the 3D magnetic field using the observed photospheric magnetograms as the boundary \cite[e.g.][]{Luoni2007a, Moreno-Insertis2008a}. In observations, the EUV radiation can further provide some information about the coronal loops (mostly traced by magnetic fields) around the null. Recently, \cite{LiuWei2011a} identified a magnetic null and considered it as the location for magnetic reconnection associated with a jet eruption. Using both the multi-perspective EUV observations and the magnetic field extrapolation, \cite{SunXuDong2012a} also found a jet eruption that was closely related to the null reconnection and determined the height of the null ($\sim$ 9 Mm). It is noted that although the configuration of the null has been presented in previous studies, the detailed evolution of the null-associated reconnection has rarely been seen. In fact, most magnetic nulls derived so far exist in the low corona \citep{Regnier2008a} and have a configuration that is only restricted to a small-scale volume. In this Letter, we present a rare large-scale null observation. The high cadence Atmospheric Imaging Assembly (AIA) data successfully uncover a magnetic null that exists as an X-shaped structure consisting of a trans-equatorial quadrupole active region group and the temporal evolution of the associated reconnection that relates to a jet eruption. In addition, we for the first time apply the differential emission measure (DEM) analysis to such a null structure using the AIA data and derive the temperature and emission measure (EM) values, which can be compared with those deduced from numerical simulations. In Section 2, we present the data reduction and results. In Section 3, we make a summary and discussion.\par

\section{Data Reduction and Results}

\subsection{Observations of an X-shaped Structure and Associated Jet}

A trans-equatorial X-shaped structure was observed by six AIA \citep{Lemen2012a} coronal passbands (including 94 {\AA}, 131 {\AA}, 171 {\AA}, 193 {\AA}, 211 {\AA}, and 335 {\AA}) from 2013 October 5 to October 7. As shown in Figure 1(a), this configuration consists of four sets of coronal loops with loop tops converging to a central region, showing a possible X-type neutral point configuration. The structure was first obviously seen in AIA 171 {\AA} images on 2013 October 5. Afterwards, it underwent a quasi static evolution and the overall pattern kept almost unchanged in the next two days.\par

Through inspecting the AIA images, it is found that two flares had occurred in succession in the region 1 (marked by a white box in Figure 1(d)) at 14:20 UT and 14:55 UT, respectively. According to the GOES soft X-ray flux, both flares are B-class and lasted only several minutes without plasma ejection. At the AIA EUV passbands, the flares appeared as bright loops that connected the emerging flux region (see $\S$ 2.2) and the northeastern part of the quadrupole active region group. About 10 minutes after the second flare, the X-shaped structure began to collapse quickly and then a jet appeared and took off from the center of the X-shaped structure with the plasma moving from east to west (see the second column of Figure 1). From the attached AIA movies, it is clearly seen that the jet eruption lasted more than 20 minutes with a projection velocity of tens of $\mathrm{km}$ $\mathrm{s^{-1}}$. The jet may be initiated in the lower corona and is part of the disturbance that caused the destruction of the X-shaped structure. However, due to the limited observations, it is difficult to know where and how exactly the jet starts. After the jet eruption, the coronal loops began to shrink with the loop tops moving away from the center of the X-shaped structure. In the region 2 (marked by the white box in Figure 1(e) and 1(f)), a set of loops brightened in 94 {\AA} passband, implying that the plasma along the loops was reaching a high temperature since this passband has a high-temperature response during flares \citep{ODwyer2010a}. About half an hour later, the jet disappeared and the coronal loops in the central part of the X-shaped structure further shrank and formed a dark region in 171 {\AA} passband (Figure 1(c)). At about 15:34 UT, a C2.3 flare occurred in region 1 (the same location with the previous two flares, see Figure 1(d)). The shape of this flare is similar to that of previous two B-class flares. At the same time, in 94 {\AA} passband, the hot loops in region 2 became wider and brighter, and gradually developed into a cusp-shaped structure at 16:30 UT (Figure 1(f)), which is most likely to be a signature of magnetic reconnection. \par

\subsection{Three-dimensional Magnetic Configuration of the X-shaped Structure}

By checking the Helioseismic and Magnetic Imager (HMI) \citep{Schou2012a} photospheric magnetic field, we find that the X-shaped structure is associated with a large-scale quadrupole field. The evolution of the line-of-sight magnetic field shows that the most notable change is the emerging flux, which is located between the two northern magnetic poles and lasts from $\sim$ 10:00 UT to $\sim$ 18:00 UT, as shown in Figure 2. Judging from the spatial distribution, the main EUV brightenings of the three flares (see $\S$ 2.1) are mainly located around the emerging flux, which may be responsible for producing the flares by reconnecting with the pre-existing magnetic fields. Notably, the second flare occurred just several minutes before the collapse of the X-shaped structure and the jet eruption and can be speculated to be the trigger of the reconnection around the null.\par 

Furthermore, we use the vector magnetic field provided by the HMI to perform a Non-Linear Force-Free Field (NLFFF) extrapolation in a Cartesian coordinate \citep{Wheatland2000a, Wiegelmann2004a}.Since the active region group is trans-equatorial, magnetic field extrapolation with a spherical coordinate may yield more accurate results \citep{Guo2012a}. Nevertheless, doing so requires a synoptic map that however loses sufficient time resolution. We also calculate the potential field and compare it with the NLFFF. It is found that the NLFFF structure in the high corona is very close to the potential one but includes more details in the low corona. The NLFFF extrapolation is performed as follows. Firstly, we remove the $180^{\circ}$ ambiguity using the the minimum energy method \citep{Metcalf1994a, Metcalf2006a, Leka2009a}. Then, we apply a projection correction to the data and degrade the resolution to 4$\arcsec$. With the extrapolation results, we further calculate the positions of the magnetic null, spine, and fan. The method to determine the magnetic null point topology has been described in \cite{Mandrini2013a}. The spine consists of two field lines connecting to the northwestern and southeastern poles, respectively. The field lines near the spine are plotted in yellow in Figure 3. These field lines change their orientations abruptly when approaching the null and then are restricted on the fan plane. The fan-spine structure here differs from those in \cite{Masson2009a} and \cite{SunXuDong2012a} in that the spine is shorter and closed locally, whereas the fan connects to further away. According to the magnetic field strength, we can easily identify the magnetic null, which is a negative one with the field lines of the spine receding it and all those of the fan approaching it. Below and above the magnetic null, one can clearly see the magnetic field structures as outlined by the observational coronal loops and jet eruption (shown in Figure 3(a) and 3(b)). At 14:48 UT, the magnetic null is located 120 Mm above the photosphere, which is about twice as high as what \cite{Filippov1999a} derived in another case. Notably, the altitude of the null rises to 130 Mm after the jet eruption at 15:12 UT. Besides, we notice that the magnetic field lines on the fan surface are partly opened during the jet eruption (Figure 3(c) and 3(d)), which may be an explanation for the upward movement of the null in order to seek for a new equilibrium.\par 

\subsection{Plasma Properties Near the X-shaped Structure}

It is widely accepted that part of energy released through magnetic reconnection goes to the plasma thermal energy. Therefore, the thermal properties of plasma, especially the temperature and EM, are critical to determine the energy release site in magnetic reconnection. Here, we perform a differential emission measure (DEM) analysis to derive the temperature and the EM of plasma near the X-shaped structure. For this purpose, we first use ``aia$\_$prep.pro'' to align the images from six passbands and degrade the resolution to 2.4$\arcsec$ in order to reduce the impact of misalignment. Then, we calculate the DEM of each pixel in the images using the method introduced by \cite{Weber2004a} and \cite{Golub2004a}. The method uses a forward modeling approach to calculate an optimal DEM. The validation of the method can be found in \cite{ChengXin2012a}.\par

To describe the overall thermal properties of the plasma around the magnetic null, we define a total emission measure and a DEM-weighted temperature as: 
 \begin{equation} 
 EM  =  \int _{T_{\mathrm{min}}} ^{T_{\mathrm{max}}} DEM(T) \mathrm{d}T,
 \end{equation}
 and
  \begin{equation} 
 T_{\mathrm{mean}} = \frac{ \int_{T_{\mathrm{min}}}^{T{\mathrm{max}}} DEM(T) T \mathrm{d}T} {\int_{T {\mathrm{min}}}^{T{\mathrm{max}}} DEM(T) \mathrm{d}T}.    
 \end{equation}
As shown in Figure 4, the EM loci curves have a good constraint on the emission measure distribution in the temperature range of 5.7 $\le$ LogT $\le$ 7.1, which is thus used to calculate the total EM and DEM-weighted temperature. Note that, all of the calculated results refer to line-of-sight integrated DEM.\par 

From the DEM results, we find that the temperature and EM around the null increased significantly after the jet eruption (Figure 4). Before the jet, the temperature around the magnetic null is about 2 MK, which is somewhat lower than that of the surrounding loops (see also \citealt{Filippov1999a}). However, the plasma heating took place around the central part of the X-shaped structure when the jet erupted. At first, there just appeared several small hot regions around the null. But shortly afterwards, these small hot regions grew and merged into a large one, as shown in Figure 4(b). This hot region looks like a long stripe extending in the northeast-southwest direction, indicating that the heating also happens to the surrounding plasma around the null. Meanwhile, we find that the hot region is of high EM as shown in Figure 4(c). The increase of EM may be a result of the plasma evaporation from the chromosphere \citep{Doschek1980a, Feldman1980a}, which is a dynamic response of chromospheric heating by heat conduction or possibly non-thermal particles.\par
 
To further analyze the temperature structure of the plasma in the hot region, we choose a small area with the size of 2.4 $\arcsec$ (centered on the red cross in Figure 4) and use the mean digital number (DN) values inside it to calculate the EM distributions at 15:01 UT and 15:36 UT. The results are displayed in Figure 4(d) and 4(e), where the red lines and the black dashed lines represent the optimal EM distributions and 100 Monte Carlo simulations, respectively. The latter can be used as an estimation of the DEM inversion errors. One can see that, before the jet, most plasma lies in a temperature range of 1.0 MK to 3.1 MK with the peak EM at 1.5 MK. While, after the jet eruption, the EM of the plasma seems to evolve into a two-component distribution, with a higher temperature component appearing at $\sim$ 6.3 MK. Moreover, we note that the peak of the lower temperature component is still located at 1.5 MK with the EM undergoing a slight decline compared to that at 15:01 UT. The decrease of the plasma in lower temperatures may reflect a fact that part of plasma is heated up to higher temperatures. Consequently, the temperature and total EM rise from 2.3 MK and $1.2\times10^{27}$ $\mathrm{cm}^{-5}$ at 15:01 UT to 5.4 MK and $3.7\times10^{27}$ $\mathrm{cm}^{-5}$ at 15:36 UT, respectively.\par

\section{SUMMARY AND DISCUSSION}

In this Letter, we investigate an X-shaped structure and the associated jet eruption above a trans-equatorial quadrupole active region group. The X-shaped structure stably existed for days and was clearly observed by all the AIA coronal passbands (especially 171 {\AA}). However, due to the continuous emerging flux in the northern part of the quadrupole active region group, two flares occurred and the second one may cause the collapse of the X-shaped structure, which resulted in the immediate shrinkage of the northern loops and the formation of a hot cusp-shaped structure, indicating that magnetic reconnection took place around the null. The collapse was also accompanied with a jet eruption. After the jet, the X-shaped structure began to disappear and a dimming region subsequently appeared in 171 {\AA} passband.\par
 
Using the NLFFF extrapolation, we reconstruct the three-dimensional magnetic configuration of the trans-equatorial quadrupole active region group, which reveals a magnetic null that exists in the center of the X-shaped structure and the associated spine and fan fields. The extrapolated magnetic fields show a good consistency with the AIA observations. The four sets of coronal loops are analogous to the four sets of magnetic fields connecting the quadrupole footpoints anchored in a relative low layer. The central part of the X-shaped structure is believed to correspond to the magnetic null. Along the spines, a set of open field lines constitute an inverted Y structure, where the heated plasma can move outward and form the jet eruption.\par

Through the DEM analysis, we find that both the temperature and EM of the plasma around the magnetic null increase during the jet eruption. It strongly suggests that magnetic reconnection is working there and responsible for the plasma heating. We notice that the initial hot region caused by the plasma heating around the null has the same orientation as the fan surface. Furthermore, by checking the extrapolated magnetic field, it is found that the connectivity of the magnetic field around the null has changed significantly. There is more flux connecting to the far-end after the jet eruption, which is likely to result from the field lines on the fan being partly opened by the reconnection. We incline to believe that the above results are evidence of the torsional fan reconnection \citep{Priest2009a}, although we cannot rule out the possibility of other forms of reconnection around the null. Since the null is located at a pretty high altitude ($\sim$ 120 Mm), the plasma there is essentially too rarefied to store enough material responsible for the high EM region. However, the plasma in the chromosphere can be heated by some means (for example, heat conduction or non-thermal particles) and move to the high corona owing to pressure imbalance to supply the plasma. \par
In conclusion, since the current event is much larger than the small-scale ones with magnetic nulls, we can successfully capture the detailed evolution of the reconnection, which first heats the plasma around the null and then causes the shrinkage of the X-shaped structure and the formation of the hot flare loops. \par

\acknowledgements The authors thank Jie Zhang for valuable discussions and suggestions, and the referee for constructive comments that helped improve the paper. SDO is a mission of NASA's Living With a Star Program. J.Q.S., X.C., Y.G., M.D.D., and Y.L. are supported by NSFC under grants 10933003, 11303016, 11373023, 11203014, and NKBRSF
under grants 2011CB811402 and 2014CB744203.


\begin{figure*} 
      \vspace{-0.0\textwidth}    
      \centerline{\hspace*{0.00\textwidth}
      \includegraphics[width=1.0\textwidth,clip=]{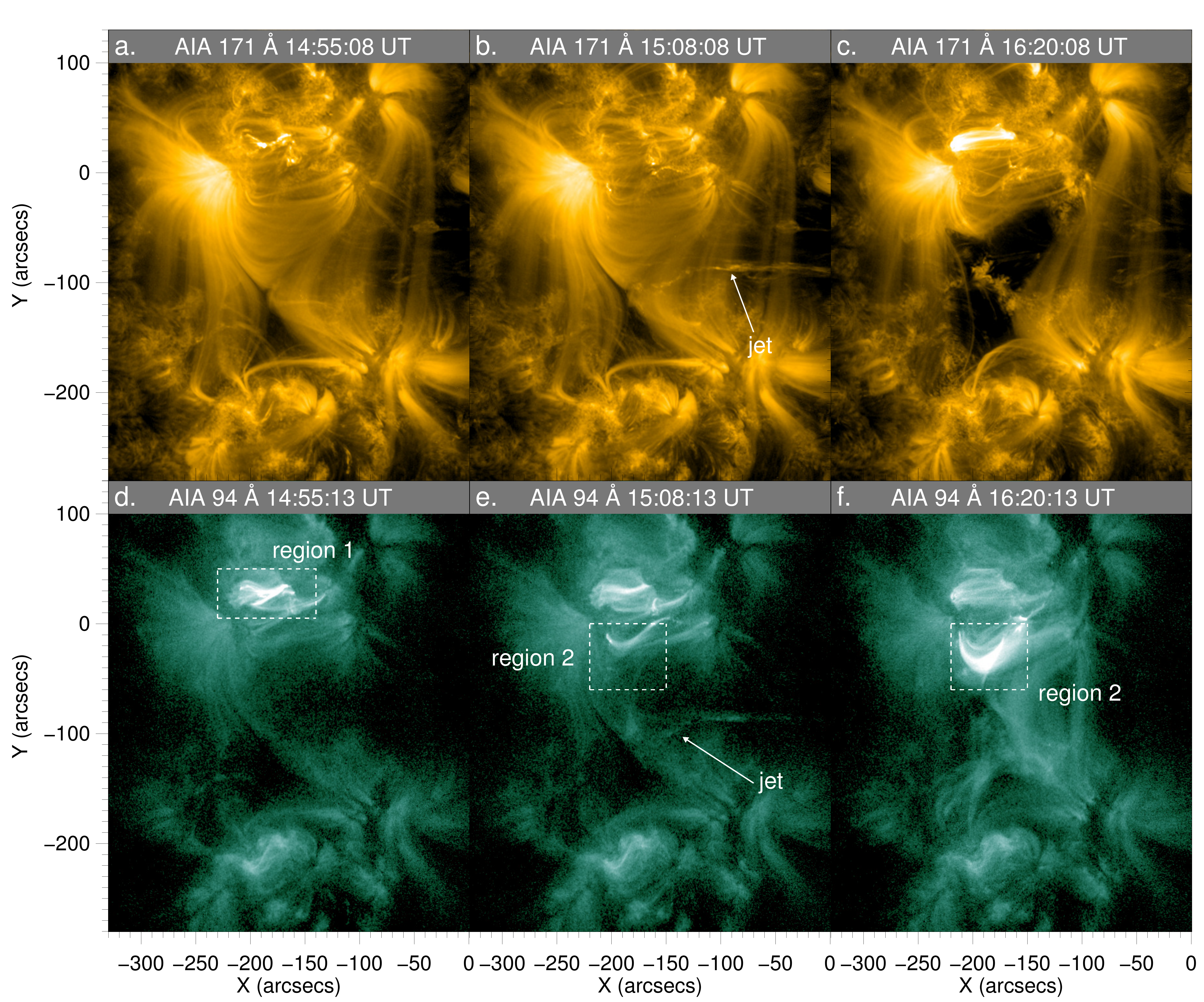}
      }
\caption{(a)--(c) AIA 171 {\AA} (Fe IX, $\sim$0.6 MK) images, (d)--(f) 94 {\AA} (Fe X, $\sim$1.1 MK; Fe XVIII, $\sim$7.1 MK) images for the X-shaped structure. } \label{f1}
(Animations are available in the online journal.)
\end{figure*}

\begin{figure*} 
     \vspace{-0.0\textwidth}    
     \centerline{\hspace*{0.00\textwidth}
               \includegraphics[width=1.0\textwidth,clip=]{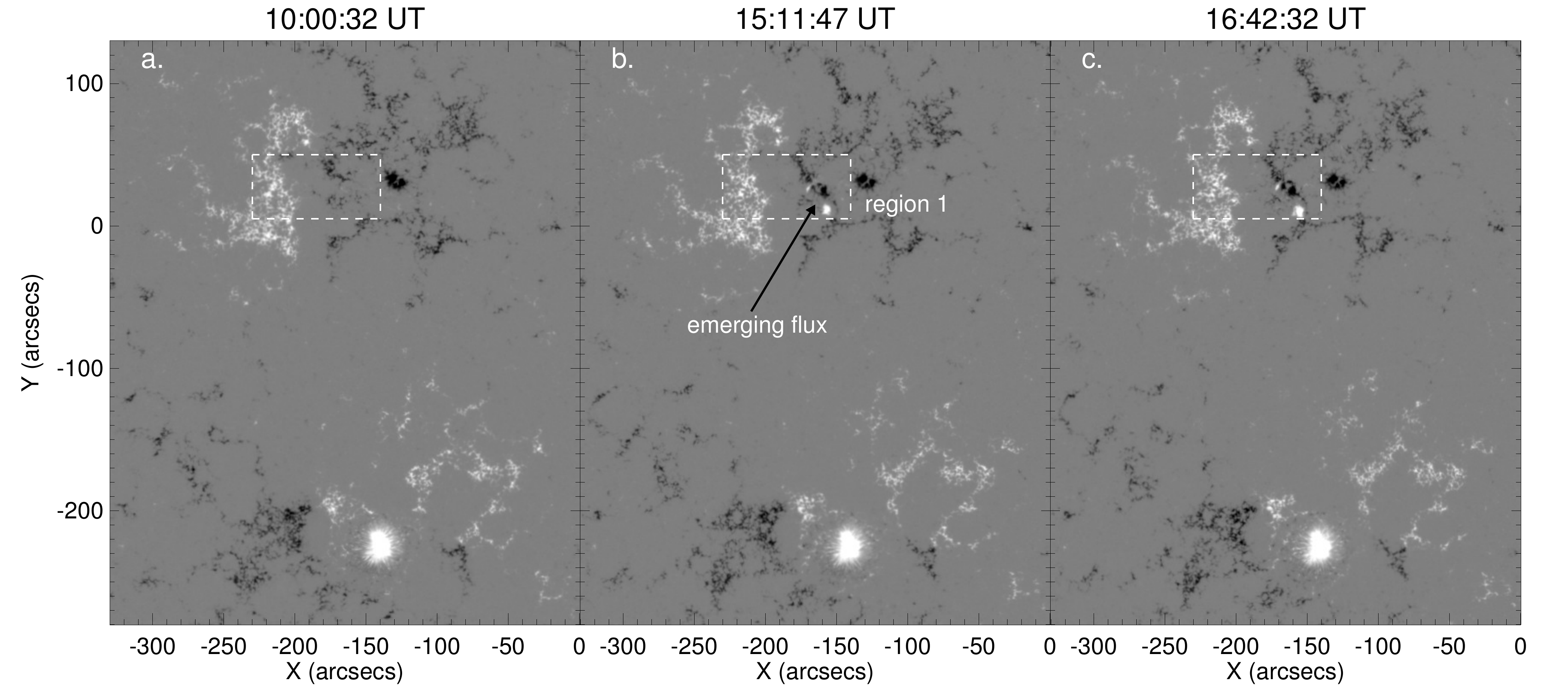}
               }
\caption{
HMI Line-of-Sight (LOS) magnetograms of the large-scale X-shaped structure at three selected instants. The white dashed boxes show the same area with region 1 in Figure 1 where two B-class flare occurred.} \label{f2}
(The animation is available in the online journal.)

\end{figure*}

\begin{figure*} 
     \vspace{-0.0\textwidth}    
     \centerline{\hspace*{0.00\textwidth}
               \includegraphics[width=1.0\textwidth,clip=]{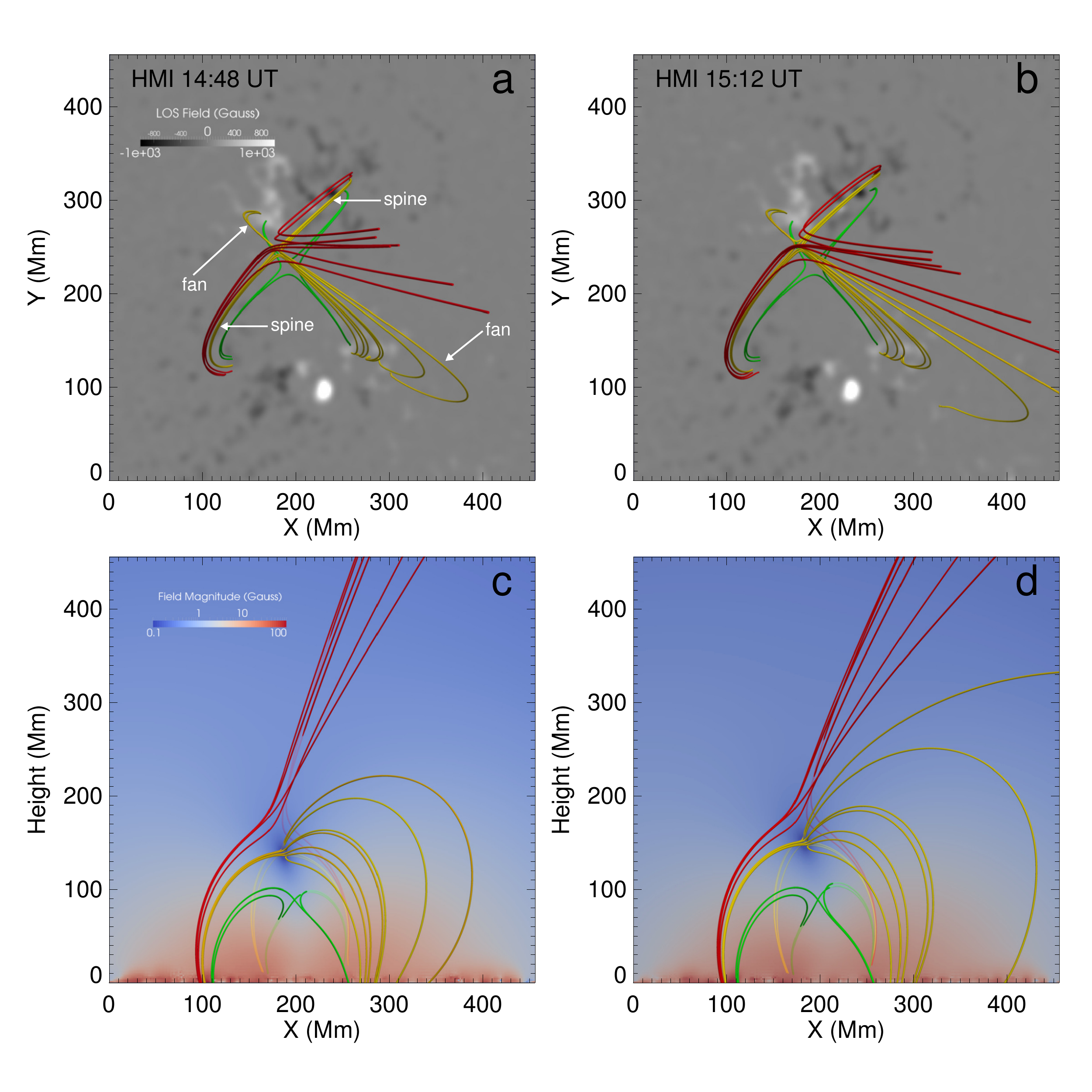}
               }
\caption{Three-dimensional NLFFF structures of the large-scale X-shaped structure. (a)--(b) Preprocessed LOS magnetograms of the active region group at 14:48 UT and 15:12 UT superposed with four sets of coronal loops and fan-spine fields above. (c)--(d) Side views of the large-scale X-shaped structure at 14:48 UT and 15:12 UT. The background represents the magnetic field strength in an x-z plane along the north-south direction containing the magnetic null . The green, yellow, and red lines denote the field lines around the particular structures discussed in the text.} \label{f3}
\end{figure*}

\begin{figure*} 
     \vspace{-0.0\textwidth}    
     \centerline{\hspace*{0.00\textwidth}
               \includegraphics[width=1.0\textwidth,clip=]{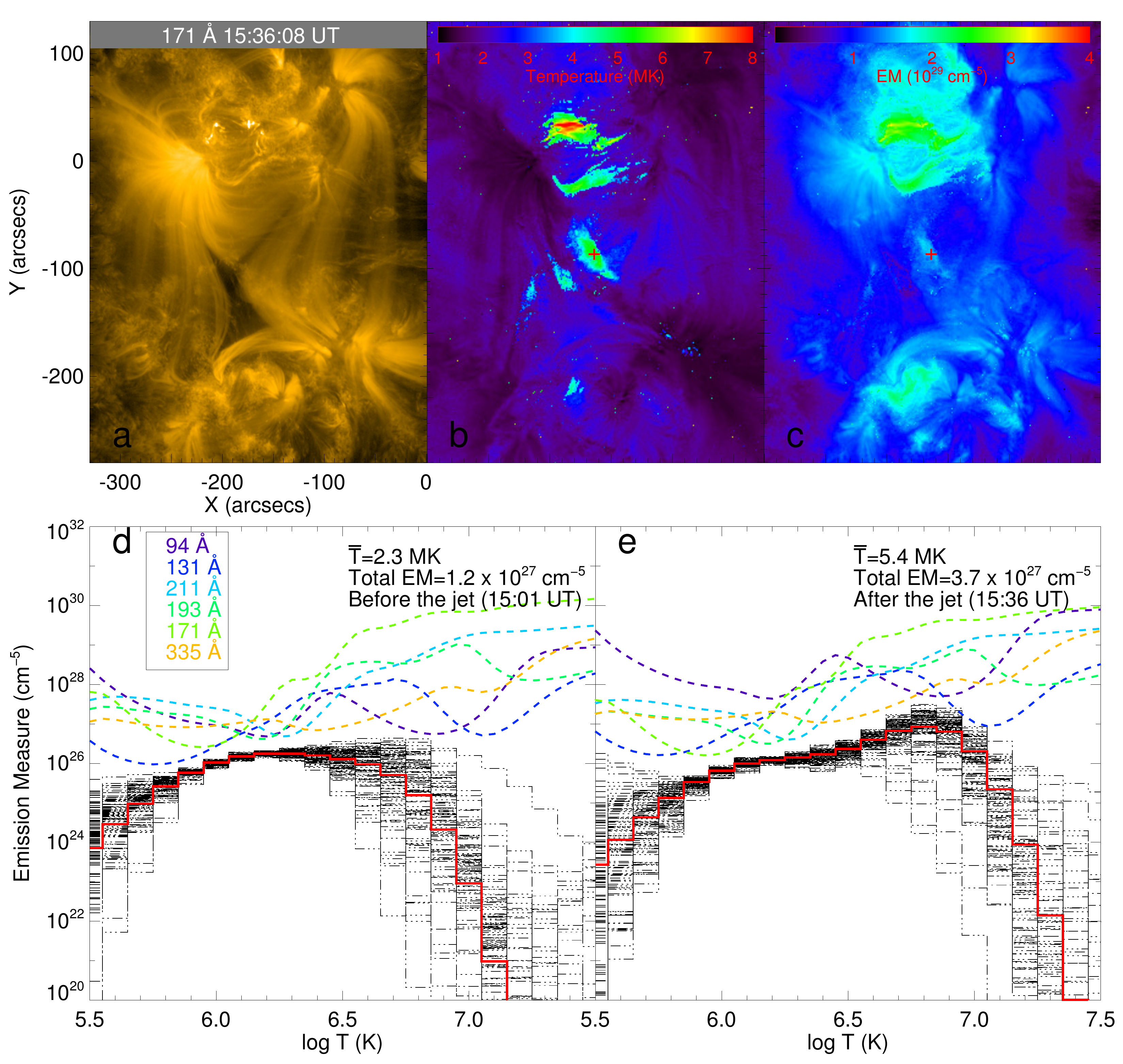}
               }
\caption{DEM results of the null point region. (a) AIA 171 {\AA} image; (b) temperature map; and (c) EM map. (d)--(e) EM distribution of the region marked by the red cross in panel (b) at 15:01 UT and 15:36 UT. The red curves are the best fitting EM distributions and the black dashed curves are 100 Monte Carlo simulations. The colored curves are the EM loci curves derived from the six AIA passbands.} \label{f4}
(Animations are available in the online journal.)
\end{figure*}

\end{document}